\documentclass[preprint,showpacs,preprintnumbers,amsmath,amssymb,nofootinbib]{revtex4}

\usepackage{graphicx}% Include figure files
\usepackage{dcolumn}% Align table columns on decimal point
\usepackage{bm}% bold math
\usepackage{longtable}
\usepackage{latexsym}
\newcommand{\ket}[1]{|{#1}\rangle}
\newcommand{\bra}[1]{\langle {#1}|}

\newcommand{\gc}[1]{\gamma\sb{#1}}
\newcommand{\vectorjc}[2]{j\sp{{#1}}\sb{V{#2}}}
\newcommand{\axialjc}[2]{j\sp{{#1}}\sb{A{#2}}}
\newcommand{\isomtx}[3]{{I}\sp{#1}({\textstyle \frac{#2}{2},\frac{#3}{2}})}

\input{epsf}
\begin{document}

%\preprint{APS/123-QED}

\title{$\Delta(1232)$ in $\pi$$N$$\rightarrow$$\pi$$\pi$$N$ reaction}

\author{Hiroyuki Kamano}
 \email{kamano@ocunp.hep.osaka-cu.ac.jp}
\author{Masaki Arima}
 \email{arima@ocunp.hep.osaka-cu.ac.jp}
\affiliation{
 Department of Physics, Osaka City University, Osaka 558-8585, Japan }

\date{\today}

\begin{abstract}
The contribution of $\Delta(1232)$ to the $\pi N\rightarrow \pi\pi N$ 
reaction is examined by making use of the
chiral reduction formula developed by Yamagishi and Zahed.
The influence of the $\pi\Delta\Delta$ and $\rho N\Delta$ interactions
on this reaction, which has not been regarded as important so far,
is considered for all channels with the initial $\pi\sp{\pm}p$ states.
The total cross sections are calculated 
to tree level for the energy up to $T\sb{\pi}=400$ MeV.
Although the $\pi\Delta\Delta$ and $\rho N\Delta$ interactions give 
a small effect on the $\pi\sp{+}p\rightarrow\pi\sp{+}\pi\sp{+}n$, 
$\pi\sp{-}p\rightarrow\pi\sp{+}\pi\sp{-}n$, and 
$\pi\sp{-}p\rightarrow\pi\sp{0}\pi\sp{0}n$ channels,
the $\pi\sp{\pm}p\rightarrow\pi\sp{\pm}\pi\sp{0}p$ channels are 
found to be sensitive to these interactions.
\end{abstract}


\maketitle

\section{\label{sec1} INTRODUCTION}%:\protect

The $\pi N\rightarrow\pi\pi N$ reaction is important 
in the analysis of the low energy $\pi N$ scattering
because of its role as a major inelastic process.  
Since the pion mass is small owing to the spontaneous 
breaking of chiral symmetry, the contribution of the 
$\pi\pi N$ channel is observed even around the $\Delta(1232)$ resonance.

Many theoretical approaches have been taken to study this reaction:
for example, the phenomenological model using effective 
Lagrangian~\cite{JM97,OV85,JDV92} 
and a series of systematic analyses based on 
the chiral perturbation theory~\cite{BKM95,BKM97,FBM00}. 
Also, the several kinds of global fits of the experimental data 
have been performed~\cite{BL91,Ver95,BPS98}. 

These  studies show that 
the $\pi N\rightarrow\pi\pi N$ reaction is useful 
to extract some important aspects of hadron physics.
The $\pi\pi$ scattering length is obtained by 
using the $\pi N\rightarrow\pi\pi N$ data.
Furthermore valuable information taken from 
the $\pi N\rightarrow\pi\pi N$ reaction benefits the 
study of baryon resonances
which have a considerable influence on this reaction.
In particular, the contribution of $\Delta(1232)$ 
is remarkable around the threshold region.

Because $\Delta$(1232) contributes to 
the $\pi N\rightarrow\pi\pi N$ reaction through three typical interactions, 
i.e., the $\pi N\Delta$, $\pi\Delta\Delta$, and $\rho N\Delta$ 
interactions, the information of these interactions can 
be accessible through the analysis of this reaction.
However the $\pi\Delta\Delta$ and $\rho N\Delta$ interactions 
have not been taken seriously in the theoretical analyses so far.
These interactions are not included in Ref.~\cite{JM97},
and their contributions to the 
$\pi\sp{-}p\rightarrow\pi\sp{+}\pi\sp{-}n$ channel
are shown to be negligible in Refs.~\cite{OV85,JDV92}.
On the other hand, many types of interactions including
the $\pi\Delta\Delta$ and $\rho N\Delta$ interactions 
are considered in the global fits~\cite{Ver95,BPS98},
but their individual roles in each channel of 
the $\pi N\rightarrow\pi\pi N$ reaction are not discussed.
The $\pi\Delta\Delta$ and $\rho N\Delta$ interactions still remain 
controversial in contrast to the well known $\pi N\Delta$ interaction.

Chiral symmetry is a key to systematic understanding of 
the pion related reactions including the $\pi N\rightarrow\pi\pi N$ reaction.
This symmetry has been accepted as a fundamental concept
in the hadron physics due to the successes of
the various low energy theorems~\cite{Current} and the 
chiral perturbation theory in the threshold region~\cite{Mei93,Eck95}.
These studies are based on spontaneous broken chiral symmetry,
where the pions are considered as the Nambu-Goldstone bosons
realized by this symmetry breaking.

Recently, a general framework analyzing hadron reactions 
has been developed by Yamagishi and Zahed~\cite{YZ96}
on the basis of chiral symmetry.
This framework introduces a new type of reduction formula, 
i.e. the chiral reduction formula,
which manifests a requirement of chiral symmetry 
satisfied by the invariant amplitudes
not only for the threshold region but for the resonance region.
It is worth noting that the chiral reduction formula
offers the Ward identity satisfied by the quantum amplitudes without
relying on any model or expansion scheme at the beginning.
This identity allows us to take a flexible and consistent view,
which is free from restrictions given by specific model, 
of the theoretical approach to resonances~\cite{SYZ97,LYZ99}.

In this paper, applying the chiral reduction formula to 
the invariant amplitude of the $\pi N\rightarrow \pi \pi N$ reaction, 
we try to clarify the contribution of  $\Delta(1232)$ in this reaction
to tree level. 
We particularly discuss  the $\pi\Delta\Delta$ and $\rho N\Delta$ 
interactions in all channels with the initial $\pi\sp{\pm}p$ state.
We assume the Rarita-Schwinger field for $\Delta(1232)$
without taking account of its internal structure.
We do not aim to fix their coupling constants but 
to clarify their importance in the $\pi N\rightarrow\pi\pi N$ reaction. 
%Our aim is not to fix their coupling constants but 
%to show those sensitivity to the $\pi N\rightarrow\pi\pi N$ reaction 
%in the exact framework of chiral symmetry.
%
Our tree level calculation is enough to make
a qualitative discussion about the $\pi\Delta\Delta$ and $\rho N\Delta$ 
interactions because of the smallness of loop collections
in this reaction as shown in ref.~\cite{FBM00} .

In Sec.~II, we show the total cross section of the
$\pi N\rightarrow\pi\pi N$ reaction.
We take into account the isospin symmetry breaking
when we compare our numerical results with the experimental values.
We explain the chiral reduction of 
the $\pi N\rightarrow \pi\pi N$ reaction in Sec.~III, 
and we show the form factors of the current matrix elements 
appeared due to the reduction of the invariant amplitude in Sec.~IV.
Our numerical results are presented in Sec.~V 
and we discuss the contributions of $\Delta(1232)$ in this reaction. 
The summaries are given in Sec.~VI.
In the Appendix, we show the phenomenological Lagrangians which are 
necessary to evaluate the form factors appeared in the current 
matrix elements, and give a brief explanation 
of the Rarita-Schwinger field.
\section{TOTAL CROSS SECTION}
\begin{figure}[ht]
\includegraphics[width=7cm]{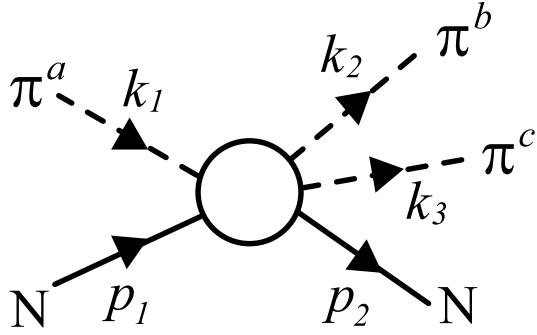}
\caption{ 
The $\pi N\rightarrow\pi\pi N$ reaction. 
Each pion has the isospin index $(a,b,c)$ and 
the four-momentum $k\sb{i}$ $(i=1,2,3)$,
and each nucleon has the four-momentum $p\sb{j}$ $(j=1,2)$.
}
\label{fig1}
\end{figure}

We consider the total cross section of the $\pi N\rightarrow \pi\pi N$ reaction 
in the isospin symmetric limit with the averaged values $m\sb{N}=939$ MeV 
and $m\sb{\pi}=138$ MeV for the nucleon and the pion masses, respectively
(see Fig.~\ref{fig1}).
The isospin symmetric total cross section is given by 
\begin{eqnarray}
\sigma\sb{\rm sym}&=&
\frac{{\cal B}}{2\sqrt{(s-m\sp{2}\sb{+})(s-m\sp{2}\sb{-})}}
\int
\frac{d\sp{3}p\sb{2}}{(2\pi)\sp{3}2E\sb{2}}
\frac{d\sp{3}k\sb{2}}{(2\pi)\sp{3}2\omega\sb{2}}
\frac{d\sp{3}k\sb{3}}{(2\pi)\sp{3}2\omega\sb{3}}\nonumber\\
&&\times(2\pi)\sp{4}\delta\sp{(4)}(p\sb{1}+k\sb{1}-p\sb{2}-k\sb{2}-k\sb{3})
\overline{|{\cal T}|\sp{2}},\label{eq:1}
\end{eqnarray}
where $s=(p\sb{1}+k\sb{1})\sp{2}$ and $m\sb{\pm}=m\sb{N}\pm m\sb{\pi}$.
The energy of outgoing nucleon and pions are denoted as
$E\sb{2}$ and $\omega\sb{i}\ (i=2,3)$, respectively. 
The Bose factor ${\cal B}$ is equal to 1/2 
if the outgoing pions are identical, otherwise this factor is 1. 
The invariant amplitude is denoted as ${\cal T}$,
for which the incoming (outgoing) nucleon spin average (sum) is taken, 
$\overline{|{\cal T}|\sp{2}}=(1/2)\sum\sb{spins}|{\cal T}|\sp{2}$.

We will show the numerical result as a function of 
the incoming pion kinetic energy in the laboratory system, $T\sb{\pi}$,
which incorporates the isospin symmetry breaking observed in  
the mass difference in each reaction channel. 
The invariant amplitude is sensitive to this difference through
the pion kinetic energy of each channel near the threshold.
We consider this symmetry breaking 
by shifting the value of the kinetic energy from 
the isospin symmetric threshold to the physical threshold~\cite{BKM97}. 
The total cross section $\sigma\sb{{\rm expt}}$ 
measured by the experiments is related to 
Eq.~(\ref{eq:1}) by 
\begin{equation}
\sigma\sb{{\rm expt}}(T\sb{\pi})=
\sigma\sb{\rm sym}({\bar T}\sb{\pi}),\label{eq:2}
\end{equation}
where $\bar{T}\sb{\pi}$ is the isospin symmetric  
pion kinetic energy and $\delta T\sb{\pi}=T\sb{\pi}-{\bar T}\sb{\pi}$.
In Table \ref{tab1}, we summarize 
the value of $\delta T\sb{\pi}$ in each reaction channel.
\begin{table}[ht]
\caption{
The values of $\delta T\sb{\pi}$ in each reaction channel.
}
\label{tab1}
\begin{ruledtabular}
\begin{tabular}{cc}
channel & $\delta T\sb{\pi}$ MeV\\ \hline
$\pi\sp{+}p \rightarrow \pi\sp{+}\pi\sp{+}n$ & $+3.97$\\
$\pi\sp{+}p \rightarrow \pi\sp{+}\pi\sp{0}p$ & $-3.66$\\
$\pi\sp{-}p \rightarrow \pi\sp{+}\pi\sp{-}n$ & $+3.97$\\
$\pi\sp{-}p \rightarrow \pi\sp{0}\pi\sp{0}n$ & $-7.92$\\
$\pi\sp{-}p \rightarrow \pi\sp{-}\pi\sp{0}p$ & $-3.66$\\
\end{tabular}
\end{ruledtabular}
\end{table}
\section{CHIRAL REDUCTION FOR $\pi N\rightarrow\pi\pi N$ REACTION}
In this section, we explain the chiral reduction of 
the invariant amplitude for the $\pi N\rightarrow\pi\pi N$ reaction,
\begin{equation}
(2\pi)\sp{4}\delta\sp{(4)}(p\sb{1}+k\sb{1}-p\sb{2}-k\sb{2}-k\sb{3})i{\cal T}
=
\bra{N(p\sb{2})}a\sp{c}(k\sb{3})a\sp{b}(k\sb{2})
\hat{\cal S}a\sp{a \dag}(k\sb{1})\ket{N(p\sb{1})}|\sb{\phi=0},
\label{eq:3}
\end{equation}
where $a\sp{a}(k)$ [$a\sp{a\dag}(k)$] is  
an annihilation (creation) operator of the pion 
with the isospin component $a$ and the four-momentum $k$. 
$\hat{\cal S}=\hat{\cal S}[\phi]$ is the extended $S$ matrix operator which 
is a functional of $\phi=(a,v,s,J)$; 
the axial vector, vector, scalar, and pseudoscalar 
$c$-number external fields~\cite{YZ96}.
At $\phi=0$, $\hat{\cal S}$ is reduced to 
the ordinary $S$ matrix operator. 
These external fields play an important role 
in the formulation of the chiral reduction formula.
Using the chiral reduction formula, we decompose 
the invariant amplitude $i{\cal T}$ as~\cite{SYZ98}
\begin{eqnarray}
i{\cal T}&=&(i{\cal T}_{\pi}+
i{\cal T}_{A}+i{\cal T}_{SA}+i{\cal T}_{VA})\nonumber\\ 
&&+(k\sb{1},a\leftrightarrow -k\sb{3},c)
+(k\sb{2},b\leftrightarrow k\sb{3},c)\nonumber\\
&&+i{\cal T}_{AAA},\label{eq:4}
\end{eqnarray}
where $(\ \leftrightarrow\ )$ represents a permutation of 
the momentum and isospin indices of the pion in the first four terms. 
Each term on the right-hand side of Eq.~(\ref{eq:4}) 
is explicitly written as
\begin{eqnarray}
i{\cal T}_\pi&=&
\frac{i}{f\sb{\pi}\sp{2}}[(k\sb{1}-k\sb{2})\sp{2}-m\sb{\pi}\sp{2}]\delta\sp{ab}
\bra{N(p_2)}\hat\pi\sp{c}(0)\ket{N(p_1)},\label{eq:5}\\
i{\cal T}_A&=&
\frac{1}{2f\sb{\pi}\sp{3}}(k_2-k_1)^\mu\delta\sp{ab}
\bra{N(p_2)}\axialjc{c}{\mu}(0)\ket{N(p_1)},\label{eq:6}\\
i{\cal T}_{SA} &=&
-i\frac{m\sb{\pi}\sp{2}}{f\sb{\pi}\sp{2}}k\sb{3}\sp{\mu}
\delta\sp{ab}\int d^4x e\sp{-i(k_1-k_2)\cdot x}
\bra{N(p_2)}T^\ast (\hat{\sigma}(x)\axialjc{c}{\mu}(0))\ket{N(p_1)},
\label{eq:7}\\
i{\cal T}_{VA} &=&
\frac{1}{2f\sb{\pi}\sp{3}}(k\sb{1}+k\sb{2})\sp{\mu}k\sb{3}\sp{\nu}
\varepsilon\sp{abe}\int d^4x e\sp{-i(k_1-k_2)\cdot x}
\bra{N(p_2)}T^\ast (\vectorjc{e}{\mu}(x)\axialjc{c}{\nu}(0))\ket{N(p_1)},
\label{eq:8}\\
i{\cal T}_{AAA} &=&
-\frac{1}{f\sb{\pi}\sp{3}}k\sb{1}\sp{\mu}k\sb{2}\sp{\nu}k\sb{3}\sp{\lambda}\nonumber\\
&&\times\int d^4x_1d^4x_2 e\sp{-ik_1\cdot x_1+ik_2\cdot x_2}
\bra{N(p_2)}T^\ast (\axialjc{a}{\mu}(x_1)\axialjc{b}{\nu}(x_2)
\axialjc{c}{\lambda}(0))\ket{N(p_1)}.\label{eq:9}
\end{eqnarray}
The pseudoscalar density $\hat{\pi}\sp{a}(x)$ is identified with
the asymptotic pion field $\pi\sb{as}(x)$ as $x\sb{0}\rightarrow\pm \infty$. 
The one-pion reduced axial current $\axialjc{a}{\mu}(x)$ is 
defined by 
$\axialjc{a}{\mu}(x)={\bf A}\sp{a}\sb{\mu}(x)
+f\sb{\pi}\partial\sb{\mu}\hat{\pi}\sp{a}(x)$ 
where ${\bf A}\sp{a}\sb{\mu}(x)$ 
is the ordinary axial current with the asymptotic form
${\bf A}\sp{a}\sb{\mu}(x)\rightarrow 
-f\sb{\pi}\partial\sb{\mu}\pi\sp{a}\sb{as}(x)+\cdots$
as $x\sb{0}\rightarrow \pm\infty$.
The vector current and the scalar density are represented by
$\vectorjc{a}{\mu}(x)$ and $\hat{\sigma}(x)$, 
respectively\footnote
{We call $\hat{\pi}\sp{a}(x)$ and $\hat{\sigma}(x)$ 
the {\textquoteleft current\textquoteright} instead of
the {\textquoteleft density\textquoteright}.}.

Equation~(\ref{eq:4}) is an exact relation 
between the invariant amplitude and the current correlations
constrained by chiral symmetry with the 
partially conserved axial-vector current (PCAC) condition, 
$\partial\sp{\mu}{\bf A}\sp{a}\sb{\mu}(x)\rightarrow
f\sb{\pi}m\sb{\pi}\sp{2}\pi\sp{a}\sb{as}(x)$ as 
$x\sb{0}\rightarrow \pm\infty$.
Thus the chiral reduction formula can naturally deal 
with the explicit breaking of chiral symmetry.
For example, ${\cal T}\sb{SA}$ is due to 
this breaking and vanishes in chiral limit $m\sb{\pi}\rightarrow0$.

\begin{figure}[ht]
  \includegraphics[width=15cm]{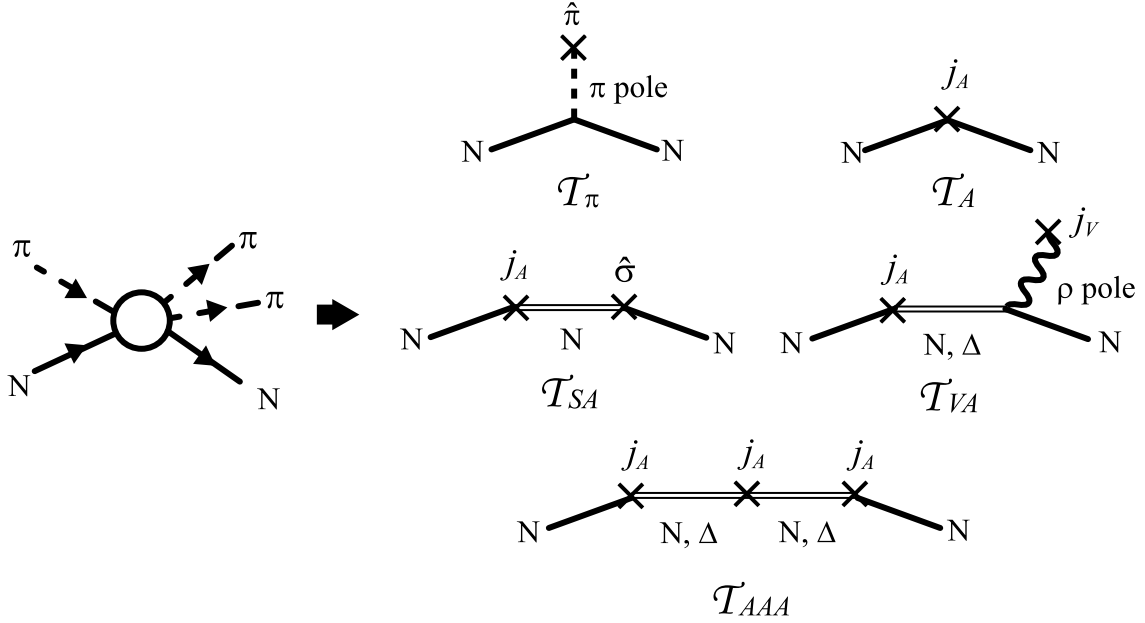}
\caption{Diagrammatical interpretation for the chiral reduction of 
$\pi N\rightarrow\pi\pi N$ reaction to tree level.
}
\label{fig2}
\end{figure}

In Fig.~\ref{fig2}, we show the diagrammatical 
interpretation of Eq.~(\ref{eq:4}) to tree level.
The solid line corresponds to the external nucleon, and 
the double line to the propagation of 
the nucleon or $\Delta(1232)$.
The cross denotes that the hadron couples to the current. 
The pion pole appears in the $\hat\pi$ current,  
while the $\axialjc{}{}$ current is free of this pole by definition.
We assume the vector meson dominance (VMD) for the 
$\vectorjc{}{}$ current, and
take into account the $\rho$ meson pole in ${\cal T}\sb{VA}$.
The $\Delta(1232)$ resonance 
does not propagate in ${\cal T}\sb{SA}$ because
the $N$-$\Delta$ transition is not brought about 
by the $\hat\sigma$ current.
\section{CURRENT MATRIX ELEMENTS}
Taking account of the Lorentz, isospin, parity and time reversal
invariances and the vector current conservation 
$\partial\sp{\mu}\vectorjc{a}{\mu}(x)=0$,
we can write the general forms of the current matrix elements 
necessary to evaluate Eqs.~(\ref{eq:5})-(\ref{eq:9}).
We summarize those matrix elements 
in Table \ref{tab2}. 
\begin{table}[ht]
\caption{
The current matrix elements necessary to evaluate each diagram
in Fig.~\ref{fig2}.
}
\label{tab2}
\begin{ruledtabular}
\begin{tabular}{ccccc}
${\cal T}\sb{\pi}$		&
${\cal T}\sb{A}$		&
${\cal T}\sb{SA}$		&
${\cal T}\sb{VA}$		&
${\cal T}\sb{AAA}$		\\ \hline
$\bra{N}\hat{\pi}\ket{N}$ 	&
$\bra{N}\axialjc{}{}\ket{N}$	&
$\bra{N}\axialjc{}{}\ket{N}$ 	&
$\bra{N}\vectorjc{}{}\ket{N}$ 	&
$\bra{N}\axialjc{}{}\ket{N}	$ \\ 
				&
				&
$\bra{N}\hat{\sigma}\ket{N}$ 	&
$\bra{\Delta}\vectorjc{}{}\ket{N}$& 
$\bra{\Delta}\axialjc{}{}\ket{N}$ \\
				&
				&
				&
$\bra{N}\axialjc{}{}\ket{N}$ 	&
$\bra{\Delta}\axialjc{}{}\ket{\Delta}$ \\
				&
				&
				&
$\bra{\Delta}\axialjc{}{}\ket{N}$&
				\\
\end{tabular}
\end{ruledtabular}
\end{table}
\begin{center}
\bf{A. Vector-isovector part}
\end{center}

The vector current matrix elements are generally written as
\begin{equation}
\bra{N(p\sp{\prime})}\vectorjc{a}{\mu}(0)\ket{N(p)}=
\bar{u}(p\sp{\prime})\left[ F\sp{N}\sb{V,1}(t)\gc{\mu}+
F\sp{N}\sb{V,2}(t)\frac{i}{2m\sb{N}}\sigma\sb{\mu\nu}q\sp{\nu}\right]
 \frac{\tau\sp{a}}{2}u(p)\label{eq:10}
\end{equation}
for the nucleon, and
\begin{eqnarray}
\bra{\Delta(p\sp{\prime})}\vectorjc{a}{\mu}(0)\ket{N(p)}&=&
\bar{U}\sp{\nu}(p\sp{\prime})\left[F\sp{N\Delta}\sb{V,1}(t)g\sb{\nu\mu}
+F\sp{N\Delta}\sb{V,2}(t)Q\sb{\nu}\gc{\mu}\right.\nonumber\\
&&\left.+F\sp{N\Delta}\sb{V,3}(t)Q\sb{\nu}Q\sb{\mu}+
iF\sp{N\Delta}\sb{V,4}(t)Q\sb{\nu}\sigma\sb{\mu\lambda}
Q\sp{\lambda}
\right]\gc{5}
\isomtx{a}{3}{1}u(p)\label{eq:11}
\end{eqnarray}
for the $N$-$\Delta$ transition, 
where $q\sp{\mu}=(p\sp{\prime}-p)\sp{\mu}$, $Q\sp{\mu}=-q\sp{\mu}$
and $t=(p\sp{\prime}-p)\sp{2}$,
$\tau\sp{a}$ is the isospin Pauli matrix and $I\sp{a}(i,j)$ is 
the $j\rightarrow i$ isospin transition $(2i+1)\times(2j+1)$ matrix.
The isoquadruplet Rarita-Schwinger vector-spinor 
and the isodoublet Dirac spinor are denoted as
$U\sp{\mu}(p)$ and $u(p)$, respectively. 
We note that $\gc{5}$ appears in Eq.~(\ref{eq:11})
because the Rarita-Schwinger field with spin-parity $(3/2)\sp{+}$
is employed for $\Delta(1232)$. 
The form factors ( $F\sp{N}\sb{V,i}(t)$ with $i=1,2$ 
and $F\sp{N\Delta}\sb{V,j}(t)$ with $j=1\sim 4$) 
are the functions of $t$.

In order to determine the $t$ dependence of the form factors to tree level, 
we make use of the current-field identity~\cite{JJ}
\begin{equation}
\vectorjc{a}{\mu}(x)=\frac{m\sb{\rho}\sp{2}}{f\sb{\rho}}\rho\sp{a}\sb{\mu}(x),
\label{eq:12}
\end{equation}
where $m\sb{\rho}$ is the $\rho$ meson mass and 
$f\sb{\rho}$ corresponds to the gauge coupling constant of 
the hidden local symmetry model for the vector mesons~\cite{Ban85,HY03}. 
Based on this identity which expresses the phenomenology of the VMD,
we write the nucleon form factors as 
\begin{eqnarray}
F\sp{N}\sb{V,1}(t)&=&
\frac{m\sb{\rho}\sp{2}}
{m\sb{\rho}\sp{2}-t-im\sb{\rho}\Gamma\sb{\rho}(t)},\label{eq:13}\\ 
F\sp{N}\sb{V,2}(t)&=&\kappa\sb{V}F\sp{N}\sb{V,1}(t)\label{eq:14},
\end{eqnarray}
where we use the standard $\rho NN$ interaction (see the Appendix)
and the universality relation $g\sb{\rho NN}=f\sb{\rho}$.
The isovector magnetic moment is denoted as $\kappa\sb{V}$.
The phenomenological width of the $\rho$ meson
is parametrized as
\begin{equation}
\Gamma\sb{\rho}(t)=\Gamma\sb{\rho}\frac{m\sb{\rho}}{\sqrt{t}}
\left(\frac{t-4m\sb{\pi}\sp{2}}
{m\sb{\rho}\sp{2}-4m\sb{\pi}\sp{2}}\right)\sp{3/2}
\theta(t-4m\sb{\pi}\sp{2}),\label{eq:15}
\end{equation}
where $\Gamma\sb{\rho}$ is 
the total width at $t=m\sb{\rho}\sp{2}$~\cite{SYZ97}.
As for the $N$-$\Delta$ transition form factors, we obtain
\begin{eqnarray}
F\sp{N\Delta}\sb{V,1}(t)&=&\frac{f\sb{\rho N\Delta}}{f\sb{\rho}}
\left(\frac{m\sb{N}+m\sb{\Delta}}{m\sb{\rho}}\right)\frac{m\sb{\rho}\sp{2}}
{m\sb{\rho}\sp{2}-t-im\sb{\rho}\Gamma\sb{\rho}(t)},\label{eq:16}\\
F\sp{N\Delta}\sb{V,2}(t)&=&\frac{1}{m\sb{N}+m\sb{\Delta}}
F\sp{N\Delta}\sb{V,1}(t),\label{eq:17}
\end{eqnarray}
where we use the $\rho N\Delta$ interaction (\ref{eq:33}). 
The $\rho N\Delta$ coupling constant is denoted as 
$f\sb{\rho N\Delta}$.

We note that the other form factors
$F\sp{N\Delta}\sb{V,3}(t)$ and $F\sp{N\Delta}\sb{V,4}(t)$ in Eq.~(\ref{eq:11})
are fixed to zero as long as we consider 
the Lagrangian (\ref{eq:33}) for the $\rho N\Delta$ interaction.
\begin{center}
\bf{B. Axial-isovector part}
\end{center}

The matrix elements of the one-pion reduced axial current 
are generally written as~\cite{KZ96, Arn79}
\begin{eqnarray}
\bra{N(p\sp{\prime})}\axialjc{a}{\mu}(0)\ket{N(p)}
&=&
\bar{u}(p\sp{\prime})
\left[
F\sp{N}\sb{A,1}(t)\gc{\mu}+
F\sp{N}\sb{A,2}(t)q\sb{\mu}
\right]
\gc{5}\frac{\tau\sp{a}}{2}u(p),
\label{eq:18}\\
\bra{\Delta(p\sp{\prime})}\axialjc{a}{\mu}(0)\ket{N(p)}
&=&
\bar{U}\sp{\nu}(p\sp{\prime})
\left[
F\sp{N\Delta}\sb{A,1}(t)g\sb{\nu\mu}
+F\sp{N\Delta}\sb{A,2}(t)Q\sb{\nu}\gc{\mu}
\right.
\nonumber\\
&&
\left.
+F\sp{N\Delta}\sb{A,3}(t)Q\sb{\nu}Q\sb{\mu}
+iF\sp{N\Delta}\sb{A,4}(t)Q\sb{\nu}\sigma\sb{\mu\lambda}Q\sp{\lambda}
\right]
\isomtx{a}{3}{1}u(p),
\label{eq:19}\\
\bra{\Delta(p\sp{\prime})}\axialjc{a}{\mu}(0)\ket{\Delta(p)}
&=&
\bar{U}\sp{\nu}(p\sp{\prime})
\left[
F\sp{\Delta}\sb{A,1}(t)g\sb{\nu\lambda}\gc{\mu}
+F\sp{\Delta}\sb{A,2}(t)g\sb{\nu\lambda}q\sb{\mu}
\right.
\nonumber\\
&&
+F\sp{\Delta}\sb{A,3}(t)(q\sb{\nu}g\sb{\mu\lambda}
+g\sb{\nu\mu}q\sb{\lambda})
\nonumber\\
&&\left.+F\sp{\Delta}\sb{A,4}(t)q\sb{\nu}\gc{\mu}q\sb{\lambda}+
F\sp{\Delta}\sb{A,5}(t)q\sb{\nu}q\sb{\mu}q\sb{\lambda}\right]
\gc{5}\isomtx{a}{3}{3}U\sp{\lambda}(p).
\label{eq:20}
\end{eqnarray}
Owing to the nature of $\Delta(1232)$ in our treatment, 
$\gc{5}$ does not appear in Eq.~(\ref{eq:19}) 
in contrast to the familiar form of the axial currents given by 
Eqs.~(\ref{eq:18}) and (\ref{eq:20}). 
Note that the pion pole does not contribute to the above form factors
by definition. 

Some of the form factors in Eqs.~(\ref{eq:18})$-$(\ref{eq:20})
are exactly related to the renormalized coupling constants for 
the pion-baryon interaction,
\begin{eqnarray}
f\sb{\pi NN}(t)&=&
\frac{m\sb{\pi}}{f\sb{\pi}}\left[\frac{1}{2}F\sp{N}\sb{A,1}(t)
+\frac{t}{4m\sb{N}}F\sp{N}\sb{A,2}(t)\right],\label{eq:21}\\
f\sb{\pi N\Delta}(t)&=&
\frac{m\sb{\pi}}{f\sb{\pi}}\left[
F\sp{N\Delta}\sb{A,1}(t)
+(m\sb{N}-m\sb{\Delta})F\sp{N\Delta}\sb{A,2}(t)
+tF\sp{N\Delta}\sb{A,3}(t)\right],\label{eq:22}\\
f\sb{\pi\Delta\Delta}(t)&=&
\frac{m\sb{\pi}}{f\sb{\pi}}\left[F\sp{\Delta}\sb{A,1}(t)+
\frac{t}{2m\sb{\Delta}}F\sp{\Delta}\sb{A,2}(t)\right],\label{eq:23}
\end{eqnarray}
where we consider the effective 
$\pi NN$, $\pi N\Delta$, and $\pi\Delta\Delta$ 
interactions (\ref{eq:29})-(\ref{eq:31}) given in the Appendix.
The other form factors  
($F\sp{N\Delta}\sb{A,4}(t)$, $F\sp{\Delta}\sb{A,3}(t)$, 
$F\sp{\Delta}\sb{A,4}(t)$ and $F\sp{\Delta}\sb{A,5}(t)$)
do not appear in our calculation.

At tree level, all the form factors are reduced to constants.
We write $F\sp{N}\sb{A,1}=g\sb{A}$ and
$F\sp{N}\sb{A,2}=-2\overline{\Delta}\sb{\pi N}/m\sb{\pi}\sp{2}$,
where $g\sb{A}$ is the axial coupling constant and 
$\overline{\Delta}\sb{\pi N}$
characterizes the discrepancy between $f\sb{\pi NN}(t=0)$ 
and $f\sb{\pi NN}(t=m\sb{\pi}\sp{2})$~\cite{SYZ98}.
%
%Due to the lack of knowledges
%for the $\pi N\Delta$ and $\pi\Delta\Delta$ interactions,
%
It is difficult to determine the differences 
in their coupling constants between $t=0$ and $t=m\sb{\pi}\sp{2}$
in the present situation of experimental data. 
Then we eliminate 
$F\sp{N\Delta}\sb{A,3}$ and $F\sp{\Delta}\sb{A,2}$ 
naturally by using the PCAC hypothesis 
$f\sb{\pi N\Delta, \pi\Delta\Delta}(m\sb{\pi}\sp{2})\simeq 
f\sb{\pi N\Delta, \pi\Delta\Delta}(0)$.
In order to determine $F\sp{\Delta}\sb{A,1}$, we use the ratio 
$R\sb{N\Delta}= f\sb{\pi\Delta\Delta}(0)/f\sb{\pi NN}(0)$ and obtain
\begin{equation}
F\sp{\Delta}\sb{A,1}=\frac{R\sb{N\Delta}}{2}g\sb{A}.\label{eq:24}
\end{equation}
In the quark model, $R\sb{N\Delta}$ becomes 4/5~\cite{BW75}.
\begin{center}
\bf{C. Scalar-isoscalar part}
\end{center}

Since $\Delta(1232)$ dose not contribute to ${\cal T}\sb{SA}$, 
we only need the nucleon matrix element for the $\hat{\sigma}$ current,
\begin{equation}
\bra{N(p\sp{\prime})}\hat{\sigma}(0)\ket{N(p)}=
S(t)\overline{u}(p\sp{\prime})u(p).\label{eq:25}
\end{equation}
According to the definition in Ref.~\cite{YZ96}, 
the form factor $S(t)$ is equal to 
$-\sigma\sb{\pi N}(t)/ f\sb{\pi}m\sb{\pi}\sp{2}$, 
where $\sigma\sb{\pi N}(t)$ is the pion-nucleon sigma term 
which becomes independent of $t$ at tree level. 
\begin{center}
\bf{D. Pseudoscalar-isovector part}
\end{center}

The nucleon matrix element of the $\hat{\pi}$ current appearing
in ${\cal T}\sb{\pi}$ is written as
\begin{equation}
\bra{N(p\sp{\prime})}\hat{\pi}\sp{a}(0)\ket{N(p)}=
P(t)\overline{u}(p\sp{\prime})i\gc{5}\tau\sp{a}u(p).\label{eq:26}
\end{equation}
This current satisfies 
\begin{equation}
(\Box +m\sb{\pi}\sp{2})\hat{\pi}\sp{a}(x)
=\frac{1}{f\sb{\pi}}\partial\sp{\mu}\axialjc{a}{\mu}(x),\label{eq:27}
\end{equation}
where the external fields are set to zero~\cite{YZ96}.
From Eqs.~(\ref{eq:18}),~(\ref{eq:27})~and~(\ref{eq:29}), 
we obtain the general relation between the form factor $P(t)$ and 
the form factors $F\sp{N}\sb{A,1}$, $F\sp{N}\sb{A,2}$
or the renormalized $\pi NN$ coupling constant,
\begin{eqnarray}
P(t)&=&
\frac{1}{m\sb{\pi}\sp{2}-t}\frac{1}{f\sb{\pi}}
(m\sb{N}F\sp{N}\sb{A,1}(t)+\frac{t}{2}F\sp{N}\sb{A,2}(t))\nonumber\\
&=&\frac{1}{m\sb{\pi}\sp{2}-t}
\left(\frac{2m\sb{N}}{m\sb{\pi}}\right)f\sb{\pi NN}(t),
\label{eq:28}
\end{eqnarray}
where the pion pole contribution is taken into account. 
\section{RESULTS AND DISCUSSIONS}
In this section we show the numerical results for the 
$\pi N\rightarrow\pi\pi N$ total cross section
below $T\sb{\pi}=400$ MeV.
We discuss the influence of 
the $\pi\Delta\Delta$ and $\rho N\Delta$ interactions
on this reaction.
The free parameters are $f\sb{\rho N\Delta}$ and $R\sb{N\Delta}$.
In Table \ref{tab3}, we summarize the value of constants in
this paper.
\begin{table}[ht]
\caption{
The value of constants in this paper.  
The mass and width of each particle or resonance are shown in MeV.
Note that $F\sp{N\Delta}\sb{A,2}$ has 
the opposite sign with $G$ in Ref.~\cite{KZ96} 
due to the difference of definition for the 
$N$-$\Delta$ transition form factors of $\axialjc{}{}$.
}
\label{tab3}
\begin{ruledtabular}
\begin{tabular}{crcl}
Masses and widths       &(MeV)  &
Parameters              &       \\ \hline
$m\sb{N}$               &939    &
$g\sb{A}$               &\ \ \ 1.265\\
$m\sb{\pi}$             &138    &
$f\sb{\rho}$            &\ \ \ 5.80\footnotemark[1]\\
$m\sb{\Delta}$          &1232   &
$\kappa\sb{V}$          &\ \ \ 3.71\\
$m\sb{\rho}$            &770    &
$F\sp{N\Delta}\sb{A,1}$ &\ \ \ 1.382\footnotemark[2]\\
$\Gamma\sb{\Delta}$     &120    &
$\overline{\Delta}\sb{\pi N}$  &$-$54\ \ \ \ \ \ \ \ \ \ \ \ \ \ \ \ MeV\footnotemark[3]\ \ \ \ \\
$\Gamma\sb{\rho}$       &149    &
$\sigma\sb{\pi N}$      &\ \ 45\ \ \ \ \ \ \ \ \ \ \ \ \ \ \ \ MeV\footnotemark[4]\\
                      &       &
$f\sb{\pi}$             &\ \ 93\ \ \ \ \ \ \ \ \ \ \ \ \ \ \ \ MeV\\
                        &       &
$F\sp{N\Delta}\sb{A,2}$ &\ $-4.235\times 10\sp{-4}$ MeV$\sp{-1}$\footnotemark[2]
\end{tabular}
\end{ruledtabular}
\footnotetext[1]{See p.33 in Ref.~\cite{HY03}}
\footnotetext[2]{Ref.~\cite{KZ96}}
\footnotetext[3]{Ref.~\cite{SYZ98}}
\footnotetext[4]{Ref.~\cite{GLS91}}
\end{table}
\begin{figure}[ht]
\includegraphics[width=7cm]{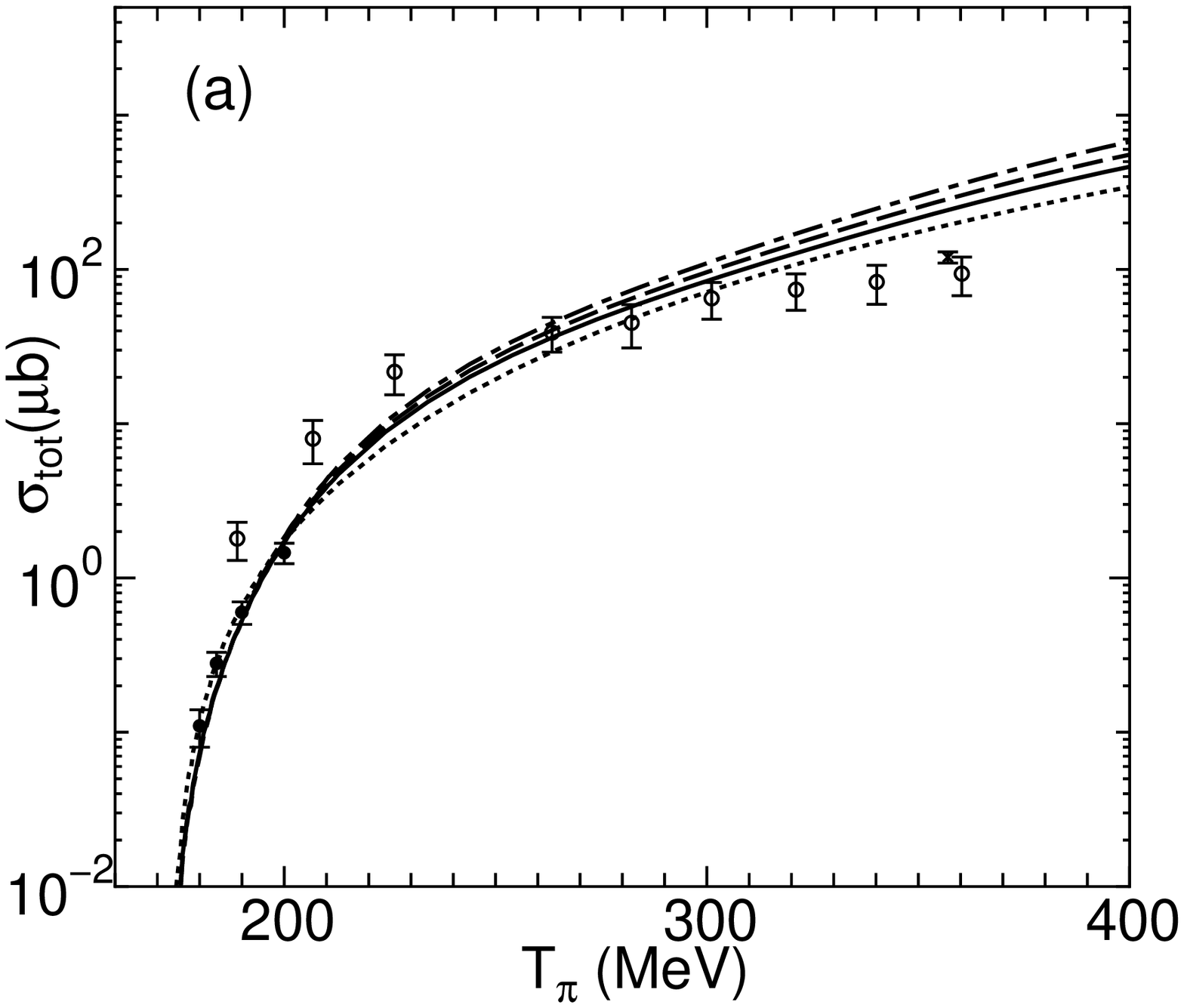} 
\includegraphics[width=7cm]{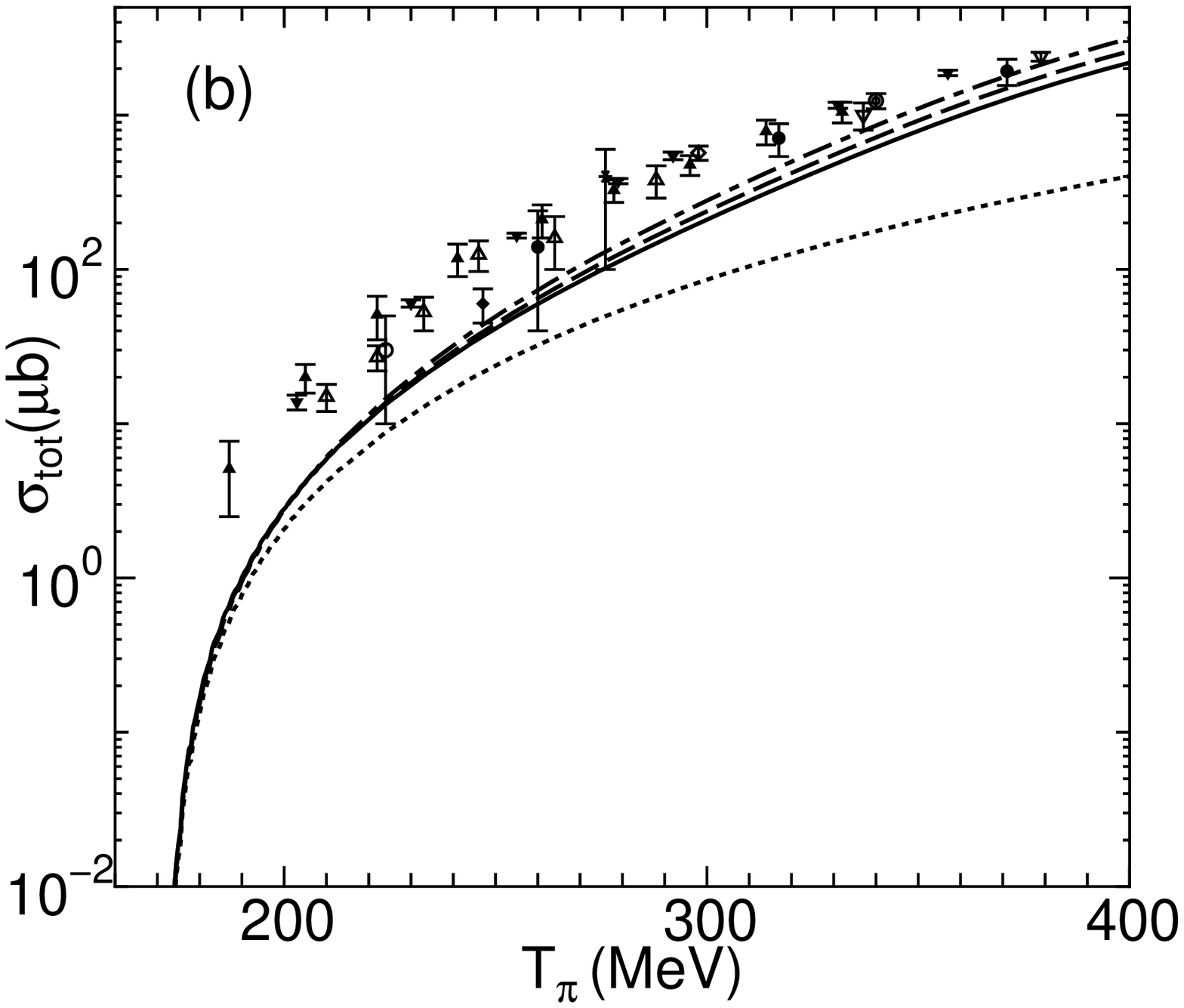}\\ 
\includegraphics[width=7cm]{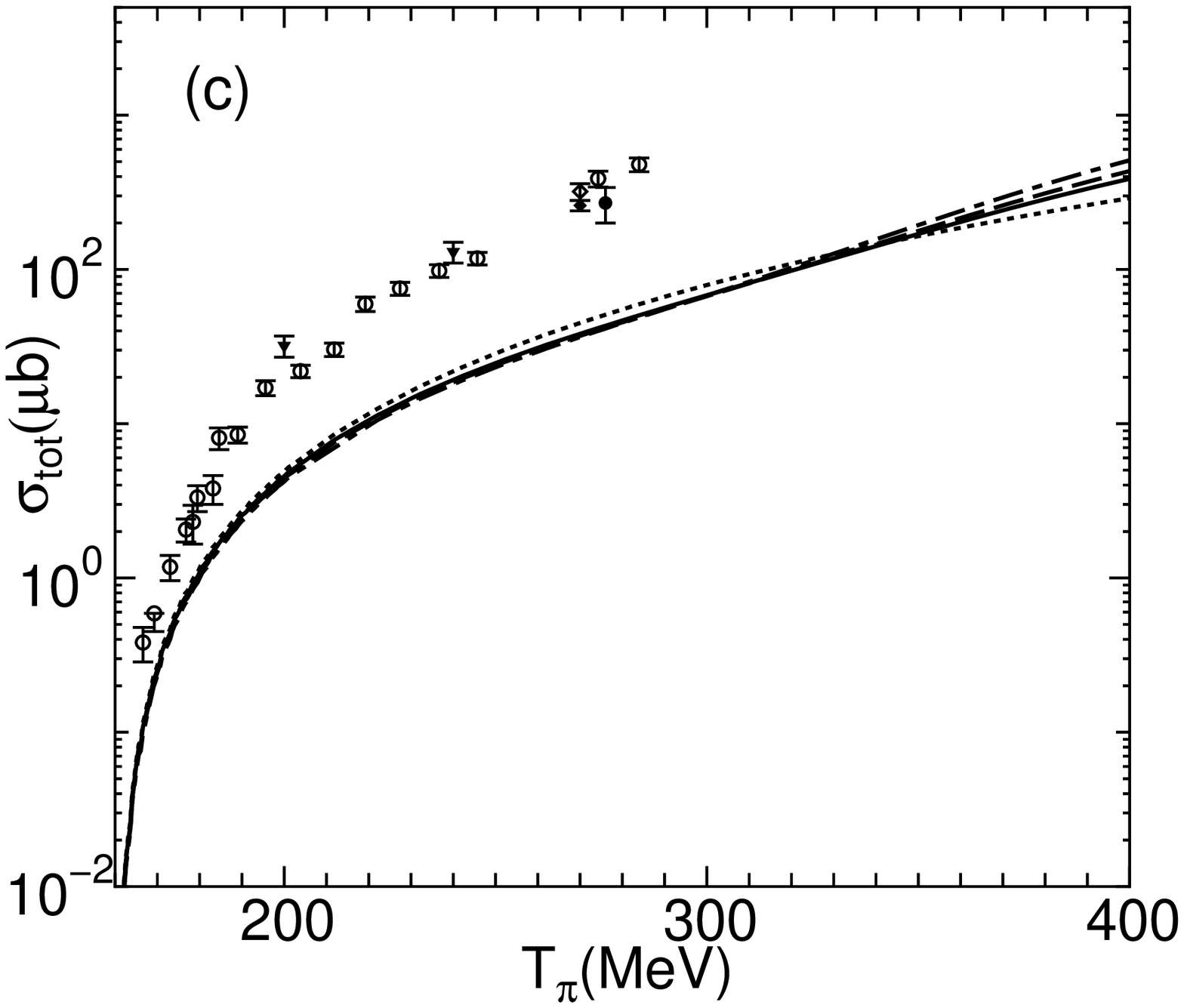}
\caption{
The dependence of the cross section on the $\pi\Delta\Delta$ interaction;
(a)~$\pi\sp{+}p\rightarrow\pi\sp{+}\pi\sp{+}n$,
(b)~$\pi\sp{-}p\rightarrow\pi\sp{+}\pi\sp{-}n$, and
(c)~$\pi\sp{-}p\rightarrow\pi\sp{0}\pi\sp{0}n$. 
It is taken $R\sb{N\Delta}=0$ for the solid line,  
$R\sb{N\Delta}=0.4$ for the dashed line, and 
$R\sb{N\Delta}=0.8$ for the dashed-dotted line. 
In addition, the dotted line is the result with the $\pi$ and $N$ only. 
The data from Refs.~\cite{Sev91,Ker90,KST62} for 
$\pi\sp{+}p\rightarrow\pi\sp{+}\pi\sp{+}n$, 
\cite{Ker89-1,Bjo80,JAS74,Bla70,SMC70,Bat65,Blo63,Dea61,Per60,Blo62}
for $\pi\sp{-}p\rightarrow\pi\sp{+}\pi\sp{-}n$, 
and \cite{Low91,Bel80,Bel78,Bun77,Kra75}
for $\pi\sp{-}p\rightarrow\pi\sp{0}\pi\sp{0}n$. 
}
\label{fig3}
\end{figure}
\begin{figure}[ht]
\includegraphics[width=7cm]{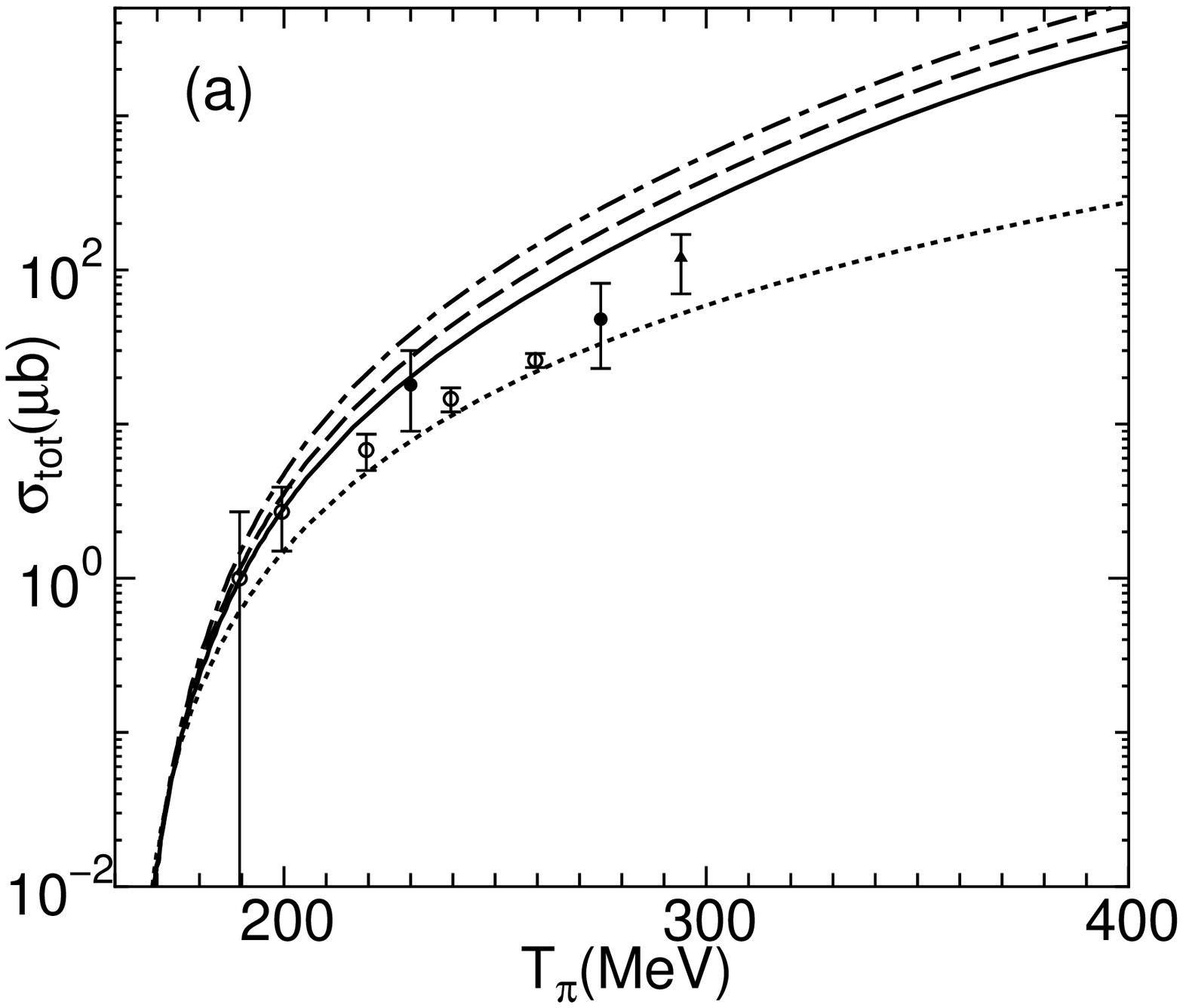} 
\includegraphics[width=7cm]{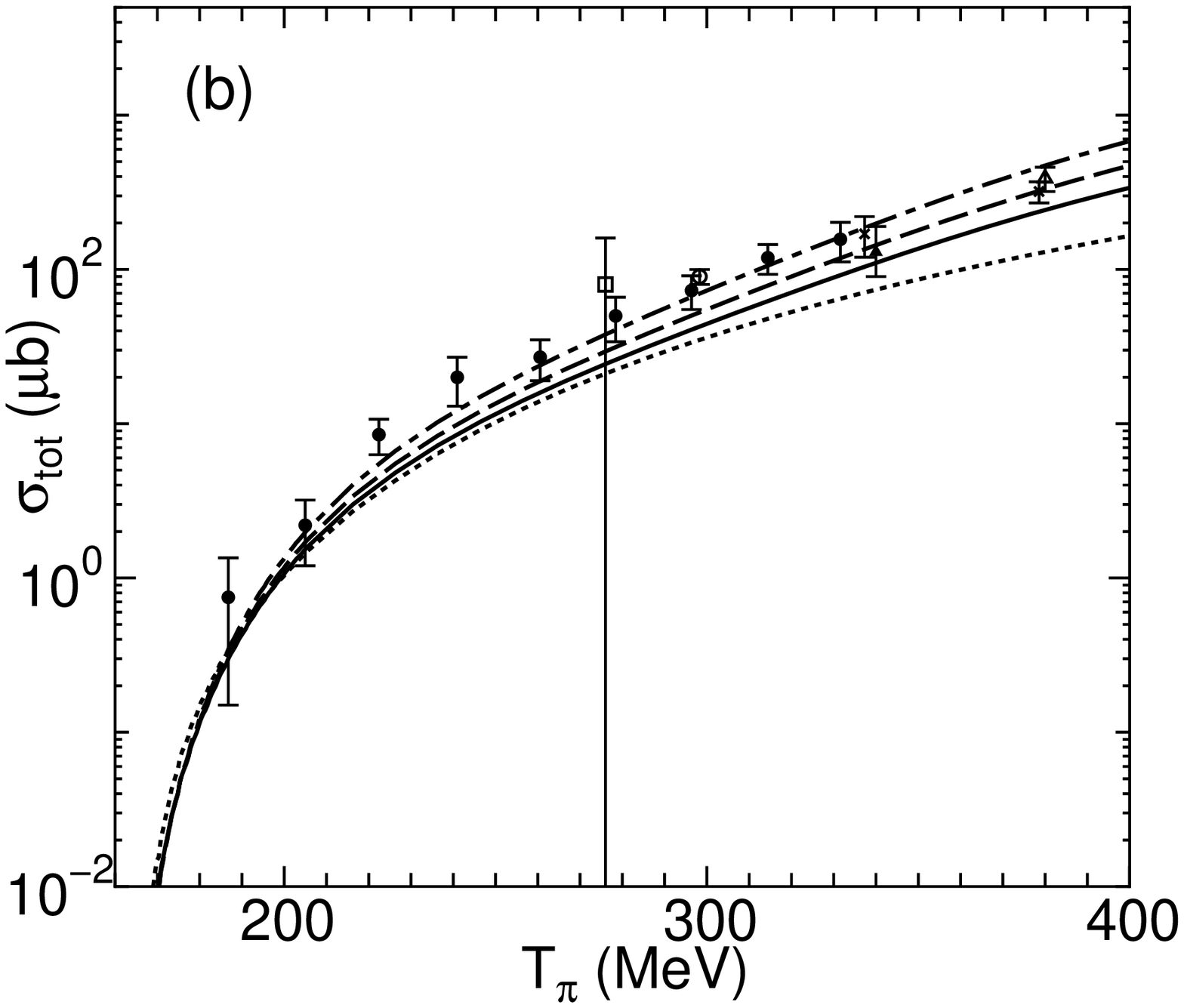}
\caption{
The dependence of the cross section on the $\pi\Delta\Delta$ interaction;
(a)~$\pi\sp{+}p\rightarrow\pi\sp{+}\pi\sp{0}p$ and
(b)~$\pi\sp{-}p\rightarrow\pi\sp{-}\pi\sp{0}p$. 
Each line is the same as that of Fig.~\ref{fig3}.
The data from Refs. \cite{Poc94,Bat75,Arm72}
for $\pi\sp{+}p\rightarrow\pi\sp{+}\pi\sp{0}p$ and
\cite{JAS74,SMC70,Blo63,Blo62,Bar61,Ker89-2}
for $\pi\sp{-}p\rightarrow\pi\sp{-}\pi\sp{0}p$.
}
\label{fig4}
\end{figure}
First we consider the $\pi\Delta\Delta$ interaction.
This interaction probes a double-$\Delta$ part of the contribution
in ${\cal T}\sb{AAA}$ when only $\Delta(1232)$ appears
as the intermediate baryon.
Figures~\ref{fig3} and \ref{fig4} show the dependence of each cross section
on the values of $R\sb{N\Delta}$. 
The $\rho N\Delta$ interaction is not included there.
Two values for $R\sb{N\Delta}$ are chosen for the calculation, 
i.e. 0.8 and 0.4.
The former is taken from the quark model result~\cite{BW75} and
the latter is selected based on the argument in Ref.~\cite{Zhu00}. 

In the $\pi\sp{+}p\rightarrow\pi\sp{+}\pi\sp{+}n$ and
$\pi\sp{-}p\rightarrow\pi\sp{+}\pi\sp{-}n$ channels,
the cross sections are increased by about 10 - 30 \% 
around $T\sb{\pi}=300$ MeV
due to the double-$\Delta$ propagation in ${\cal T}\sb{AAA}$
compared to the results with $R\sb{N\Delta}=0$. 
In $\pi\sp{-}p\rightarrow\pi\sp{0}\pi\sp{0}n$ channel,
the effect of the double-$\Delta$ propagation is less than a few percent.
In the same figure, we also display the experimental data taken from
Refs.~\cite{Sev91,Ker90,KST62,Ker89-1,Bjo80,JAS74,Bla70,SMC70,
Bat65,Blo63,Dea61,Per60,Blo62,Low91,Bel80,Bel78,Bun77,Kra75,
Poc94,Bat75,Arm72,Bar61,Ker89-2}.
As for the $\pi\sp{-}p\rightarrow\pi\sp{+}\pi\sp{-}n$ and 
$\pi\sp{-}p\rightarrow\pi\sp{0}\pi\sp{0}n$ channel, 
the disagreement between the theory and data is far beyond 
the variation in the cross section
due to the $\pi\Delta\Delta$ interaction. 
This result is consistent with that of Ref.~\cite{OV85}
for the $\pi\sp{-}p\rightarrow\pi\sp{+}\pi\sp{-}n$ channel.

In contrast with the above results,
the $\pi\sp{\pm}p\rightarrow\pi\sp{\pm}\pi\sp{0}p$ channels are
sensitive to the $\pi\Delta\Delta$ interaction (see Fig.~\ref{fig4}).
The total cross section for the
$\pi\sp{+}p\rightarrow\pi\sp{+}\pi\sp{0}p$ channel below 300 MeV
increases by about 50 \% with $R\sb{N\Delta}=0.4$, and the cross section
with $R\sb{N\Delta}=0.8$ becomes twice the result with $R\sb{N\Delta}=0$. 
As for the $\pi\sp{-}p\rightarrow\pi\sp{-}\pi\sp{0}p$ channel,
the total cross section indicates the increase by about 25 \% 
with $R\sb{N\Delta}=0.4$ and 65 \% with $R\sb{N\Delta}=0.8$ around 
$T\sb{\pi}=300$ MeV.
These two channels exhibit large influence of 
the $\pi\Delta\Delta$ interaction on the total cross section. 
However, if the fit with data is considered, 
it is still difficult for 
the $\pi\sp{+}p\rightarrow\pi\sp{+}\pi\sp{0}p$ channel
to improve the result by including the $\pi\Delta\Delta$ interaction alone.
\begin{figure}[ht]
\includegraphics[width=7cm]{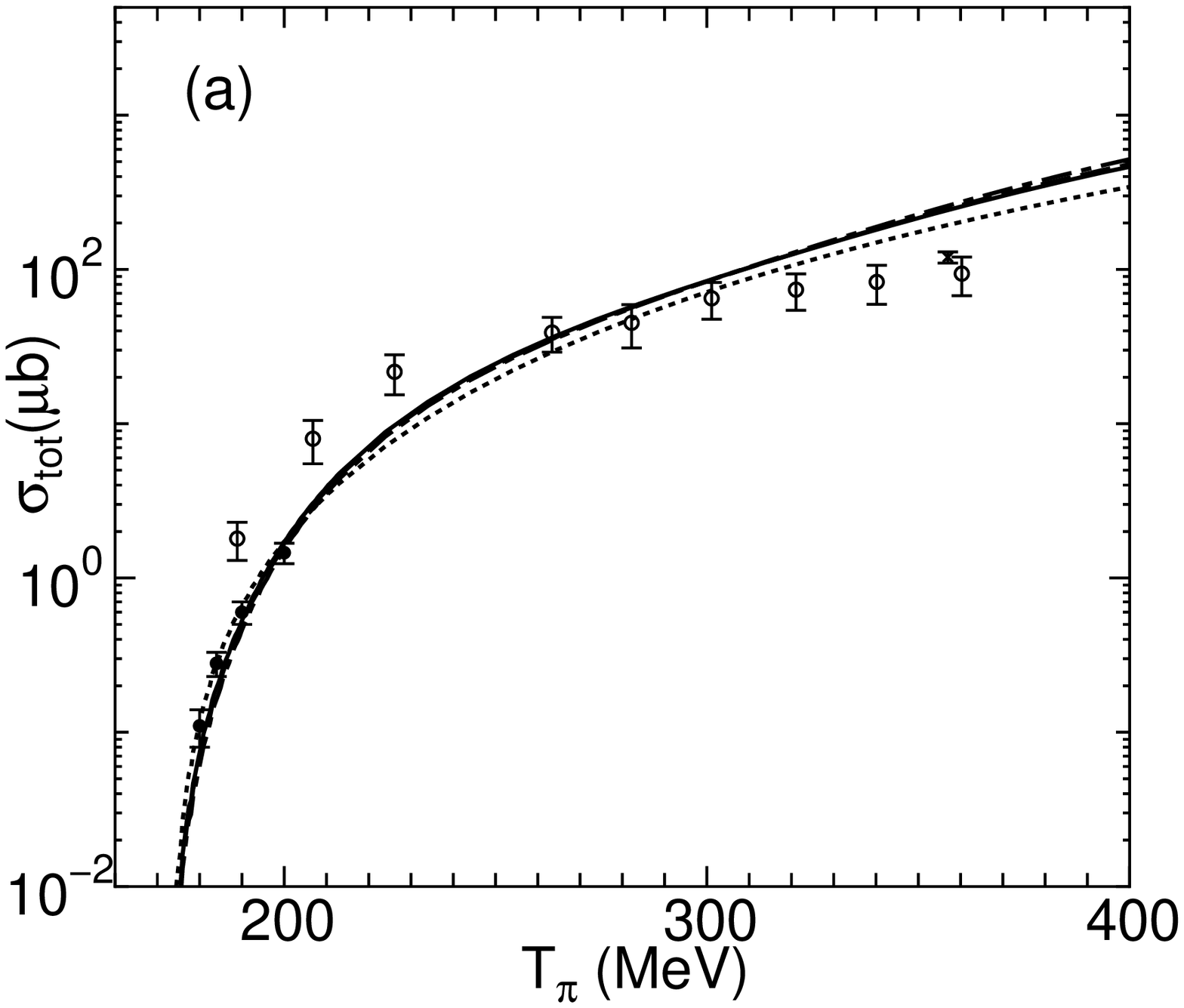} 
\includegraphics[width=7cm]{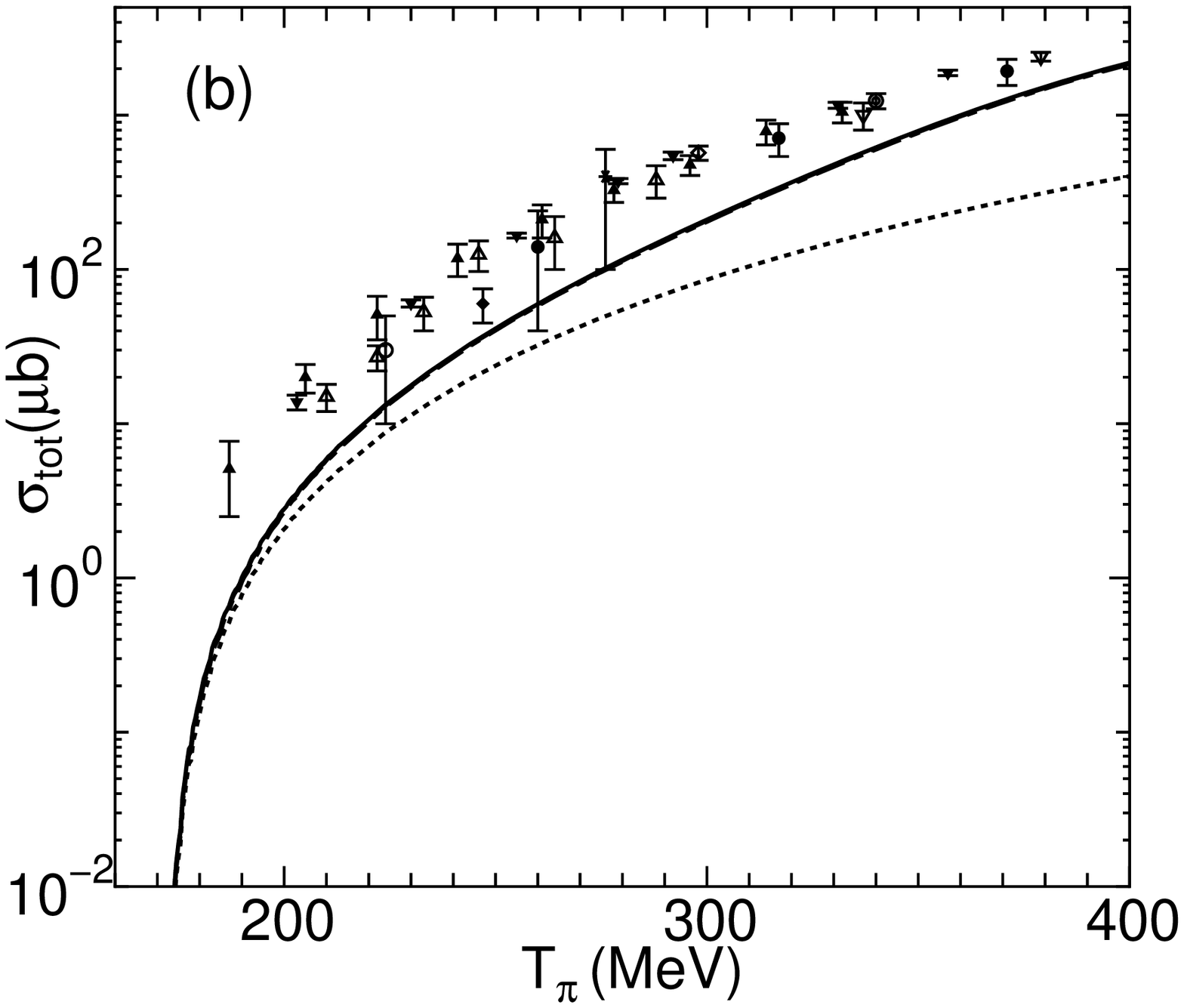}\\ 
\includegraphics[width=7cm]{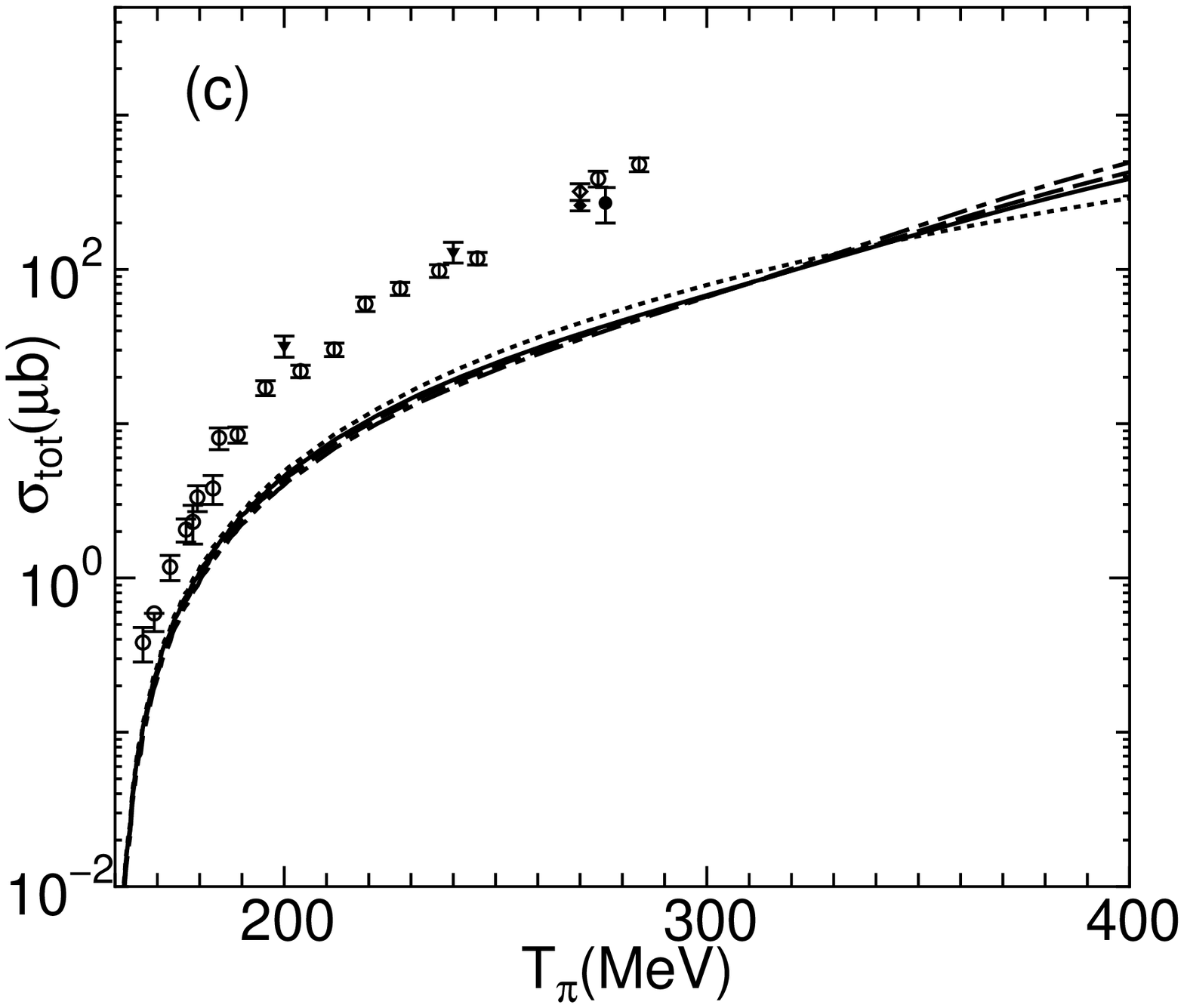}
\caption{
The dependence of the total cross section on the $\rho N\Delta$ interaction;
(a)~$\pi\sp{+}p\rightarrow\pi\sp{+}\pi\sp{+}n$,
(b)~$\pi\sp{-}p\rightarrow\pi\sp{+}\pi\sp{-}n$ and
(c)~$\pi\sp{-}p\rightarrow\pi\sp{0}\pi\sp{0}n$. 
It is taken $f\sb{\rho N\Delta}=0$ for the solid line,
$f\sb{\rho N\Delta}=3.5$ for the dashed line and
$f\sb{\rho N\Delta}=7.8$ for the dashed-dotted line.
The data and the dotted line are same as in Fig.~\ref{fig3}. 
}
\label{fig5}
\end{figure}
\begin{figure}[ht]
\includegraphics[width=7cm]{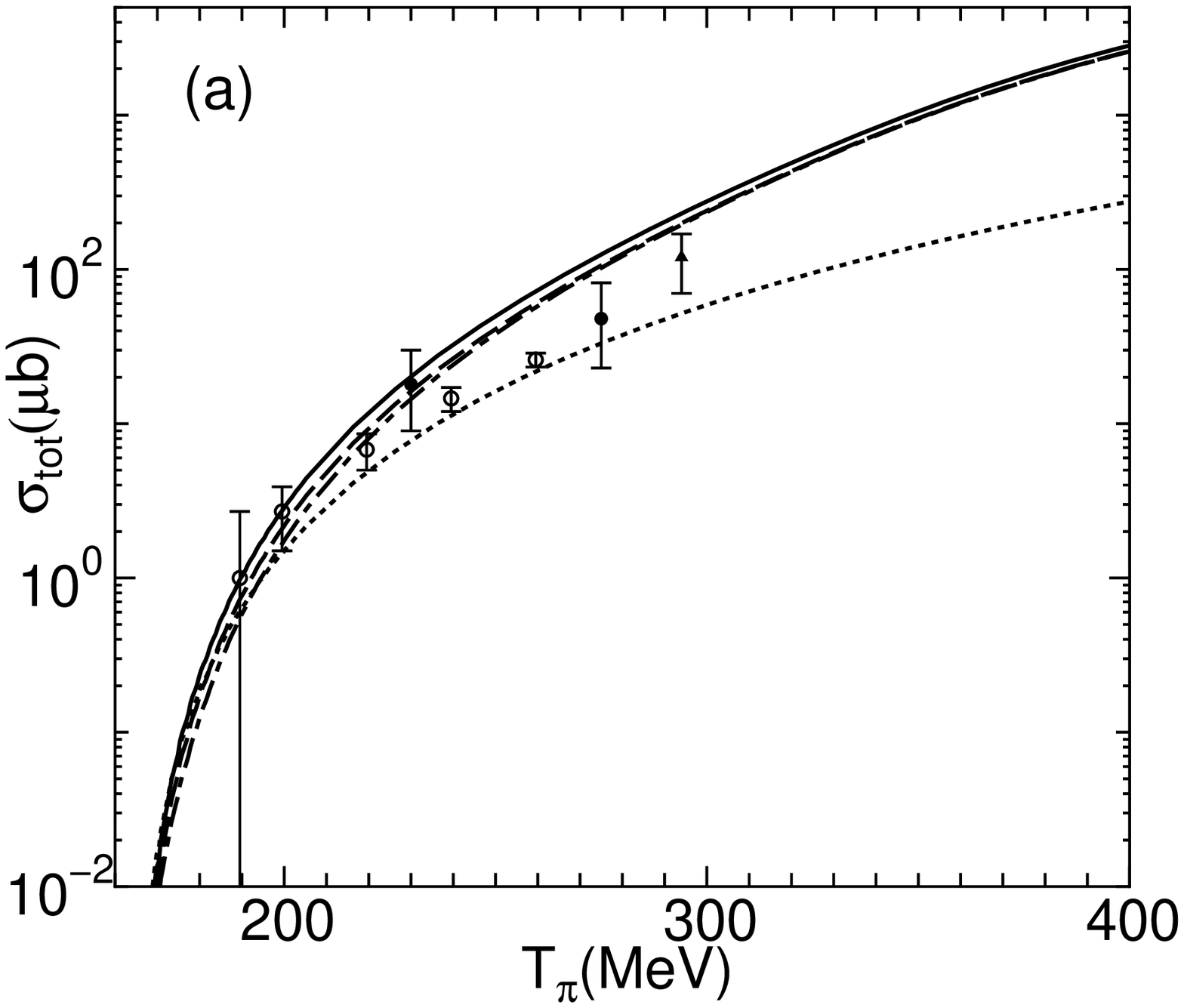} 
\includegraphics[width=7cm]{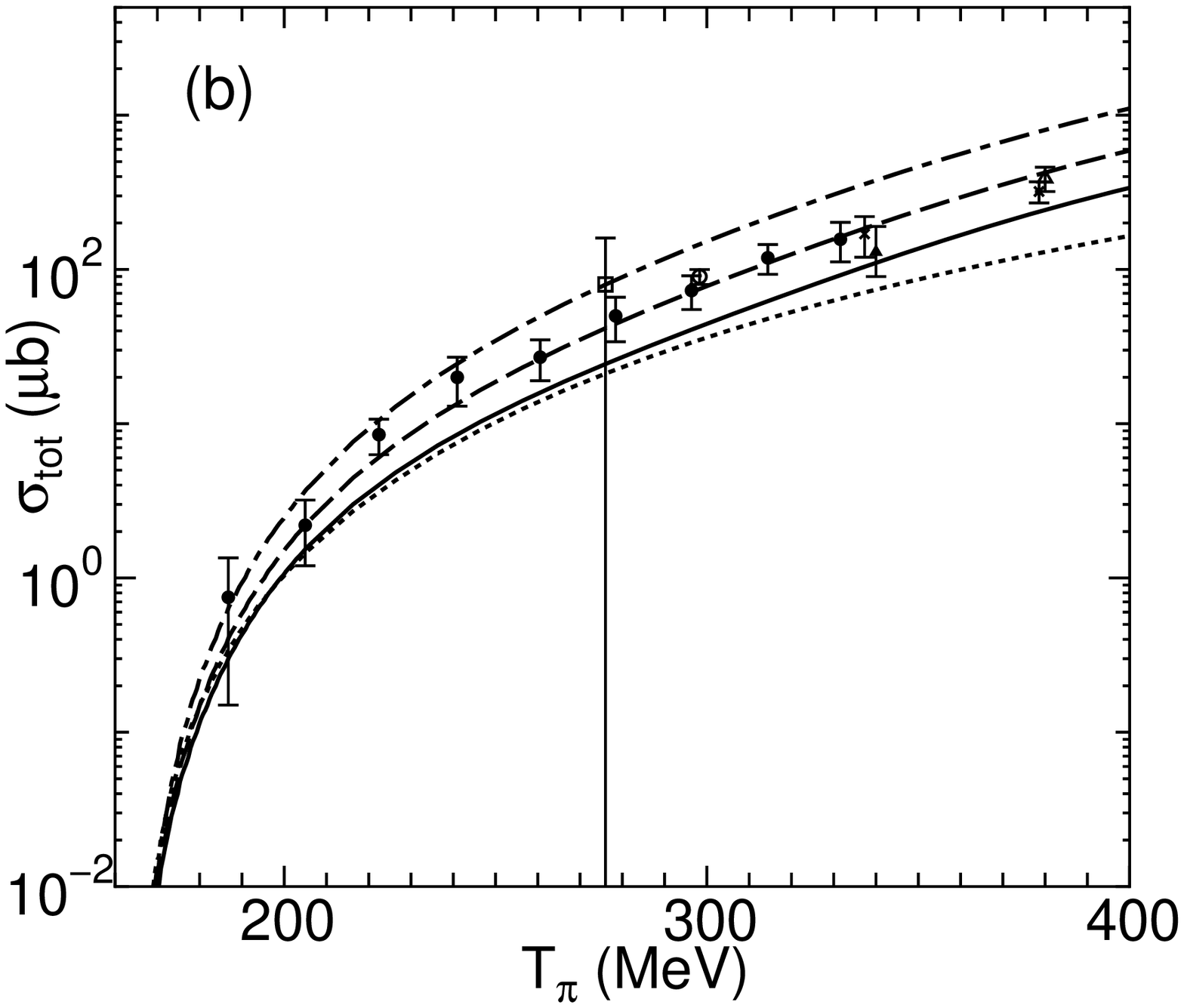}
\caption{
The dependence of the total cross section on the $\rho N\Delta$ interaction;
(a)~$\pi\sp{+}p\rightarrow\pi\sp{+}\pi\sp{0}p$ and
(b)~$\pi\sp{-}p\rightarrow\pi\sp{-}\pi\sp{0}p$. 
Each line is same as that of Fig.~\ref{fig5}.
The data is same as in Fig.~\ref{fig4}. 
}
\label{fig6}
\end{figure}
Next we show the dependence  
of each cross section on the value of $f\sb{\rho N\Delta}$
in Figs.~\ref{fig5}~and~\ref{fig6}.  
We do not include the $\pi\Delta\Delta$ interaction instead.
We take two values for $f\sb{\rho N\Delta}$, 3.5 and 7.8, 
referring to the self-consistent calculation of 
the $\rho N\Delta$ vertex function~\cite{HL94} and 
the $NN$ phase shift analysis using the meson exchange
diagram~\cite{MHE87}, respectively.

As for the three channels in Fig.~\ref{fig5}, 
the effect of the $\rho N\Delta$ interaction is negligible. 
This situation is similar to the case of Fig.~\ref{fig3}, in 
which the dependence on the $\pi\Delta\Delta$ interaction is examined.
From the results given in Figs.~\ref{fig3}~and~\ref{fig5}, 
we can say that the $\pi\sp{+}p\rightarrow\pi\sp{+}\pi\sp{+}n$
and  $\pi\sp{-}p\rightarrow\pi\sp{0}\pi\sp{0}n$ channels
are insensitive to $\Delta(1232)$ and $\rho$. 
Since the experimental data for the
$\pi\sp{+}p\rightarrow\pi\sp{+}\pi\sp{+}n$ channel
are well reproduced by the theoretical calculation, 
this channel is saturated by the reaction 
mechanisms and contains only the pion and the nucleon. 
This result is in agreement with other calculations~\cite{JM97,BKM97}.

On the other hand, the effect of the $\rho N\Delta$ interaction 
is seen in the two channels in Fig.~\ref{fig6}.
In the $\pi\sp{+}p\rightarrow\pi\sp{+}\pi\sp{0}p$ channel,  
the cross section with $f\sb{\rho N\Delta}=3.5$ and $7.8$ 
around $T\sb{\pi}=200$ MeV decreases about 30 \% 
in comparison with the cross section with $f\sb{\rho N\Delta}=0$.
In the $\pi\sp{-}p\rightarrow\pi\sp{-}\pi\sp{0}p$ channel,  
the cross section shows large increase over all values of $T\sb{\pi}$.
Many reports on the value of $f\sb{\rho N\Delta}$ are settled in the range
of $3.5\alt f\sb{\rho N\Delta} \alt 7.8$, and the 
$\pi\sp{-}p\rightarrow\pi\sp{-}\pi\sp{0}p$ channel is 
sensitive to the variation of $f\sb{\rho N\Delta}$ in this range.

Here we comment on the Roper resonance.
The experimental data for 
the $\pi\sp{-}p\rightarrow\pi\sp{+}\pi\sp{-}n$ 
and $\pi\sp{-}p\rightarrow\pi\sp{0}\pi\sp{0}n$ channels 
can not be reproduced in this paper.
It is known that this failure 
can be cured by the Roper resonance $N\sp{\ast}(1440)$ with 
$N\sp{\ast}(1440)\rightarrow N(\pi\pi)\sb{I=0}$ decay~\cite{JM97,OV85}. 
However, this resonance was not included in this paper
because we concentrate our attention on the $\Delta(1232)$ resonance.
We note that $N\sp{\ast}(1440)\rightarrow N(\pi\pi)\sb{I=0}$ contribution 
appears, at tree level, only in ${\cal T}\sb{SA}$ 
which results from explicit breaking of chiral symmetry.
Although several improvements are suggested for the treatment of 
the Roper resonance~\cite{BK02}, the role of 
this famous resonance is still in question.
\section{SUMMARY}
We have calculated the total cross section 
of the $\pi N\rightarrow\pi\pi N$ reaction up to $T\sb{\pi}=400$ MeV
and discussed the contribution of $\Delta(1232)$ to this 
reaction.
Applying the chiral reduction formula to the invariant amplitude,
we have considered the $\pi\Delta\Delta$ and $\rho N\Delta$ interactions
which have not been taken seriously so far.
Using the numerical values of the coupling constants
given in the past studies, we have shown that the 
$\pi\sp{\pm}p\rightarrow\pi\sp{\pm}\pi\sp{0}p$ channels
are sensitive to these two interactions.
If we hope to determine concrete values of these coupling constants,
we need to extend our treatment by, for example, the systematic inclusion
of the Roper resonance.

The chiral reduction formula is found to be effective to use
to analyze the $\pi N\rightarrow\pi\pi N$ reaction.
Owing to this formula we can consider the detail of each reaction  
mechanism separately from the general framework of 
pion induced reactions constrained by chiral symmetry. 
This approach has an advantage to gain deeper insight for 
the nonperturbative features of the hadron physics.
\begin{acknowledgments}
The authors acknowledge the nuclear theory group of Osaka-City University
for useful discussions.
\end{acknowledgments}
\appendix*\section{}

In this Appendix, we show the phenomenological Lagrangians 
and the $\Delta$ propagator.
Some useful relations for the isospin matrices are listed, too. 

The Lagrangians are written in the following forms,
\begin{eqnarray}
{\cal L}\sb{\pi NN}&=&
\frac{f\sb{\pi NN}}{m\sb{\pi}}
\bar{N}\gc{\mu}\gc{5}
\tau\sp{a}N\partial\sp{\mu}\pi\sp{a},\label{eq:29}\\
{\cal L}\sb{\pi N\Delta}&=&
\frac{f\sb{\pi N\Delta}}{m\sb{\pi}}\bar{\Delta}\sp{\nu}
\Theta\sb{\nu\mu}(Z)\isomtx{a}{3}{1}
N\partial\sp{\mu}\pi\sp{a}+h.c. \ ,\label{eq:30}\\
{\cal L}\sb{\pi\Delta\Delta}&=&\frac{f\sb{\pi\Delta\Delta}}{m\sb{\pi}}
\bar{\Delta}\sp{\alpha}\Theta\sb{\alpha\beta}(Z\sp{\prime})\gc{\mu}\gc{5}
\isomtx{a}{3}{3}{\Theta\sp{\beta}}\sb{\delta}
(Z\sp{\prime})\Delta\sp{\delta}\partial\sp{\mu}\pi\sp{a},\label{eq:31}\\
{\cal L}\sb{\rho NN}&=&
g\sb{\rho NN}
\bar{N}\left[\gc{\mu}\rho\sp{\mu a}-\frac{\kappa\sb{V}}{2m\sb{N}}
\sigma\sb{\mu\nu}\partial\sp{\nu}\rho\sp{\mu a}
\right]\frac{\tau\sp{a}}{2}N,\label{eq:32}\\
{\cal L}\sb{\rho N\Delta}&=&
i\frac{f\sb{\rho N\Delta}}{m\sb{\rho}}\bar{\Delta}\sp{\sigma}
\Theta\sb{\sigma\mu}(Z\sp{\prime\prime})\gc{\nu}\gc{5}
\isomtx{a}{3}{1}N
(\partial\sp{\nu}\rho\sp{\mu a}-\partial\sp{\mu}\rho\sp{\nu a})+h.c. \ ,
\label{eq:33}
\end{eqnarray}
where $(\Delta\sb{\mu})\sp{\rm T}=
(\Delta\sp{++}\sb{\mu},\Delta\sp{+}\sb{\mu},
\Delta\sp{0}\sb{\mu},\Delta\sp{-}\sb{\mu})$
and $N\sp{\rm T}=(p,n)$.
The second-rank Lorentz tensor $\Theta\sb{\mu\nu}$ is defined by 
$\Theta\sb{\mu\nu}(Y)\equiv g\sb{\mu\nu}-
\frac{1}{2}(1+2Y)\gc{\mu}\gc{\nu}$ (where we take $A=-1$ \cite{NEK71}). 
The second term of this tensor vanishes if $\Delta$ is on the mass shell
(because of $\gc{\mu}U\sp{\mu}(p)=0$), 
so that $Y$ is called the off-shell parameter. 
In this paper, we assume 
$Z=Z\sp{\prime}=Z\sp{\prime\prime}=-\frac{1}{2}$ for simplicity, that is
$\Theta\sb{\mu\nu}\rightarrow g\sb{\mu\nu}$.

The $\Delta(1232)$ propagator is 
\begin{equation}
S\sb{\mu\nu}(p)=
\frac{(\not{p}+m\sb{\Delta})}{3(p\sp{2}-m\sp{2}\sb{\Delta})}
\left[-2g\sb{\mu\nu}+\frac{2p\sb{\mu}p\sb{\nu}}{m\sb{\Delta}\sp{2}}
-i\sigma\sb{\mu\nu}
+\frac{\gc{\mu}p\sb{\nu}-\gc{\nu}p\sb{\mu}}{m\sb{\Delta}}
\right].\label{eq:34}
\end{equation}
In order to include the $\Delta(1232)$ width phenomenologically, 
we modify the denominator of the $\Delta$ propagator as
$p\sp{2}-m\sb{\Delta}\sp{2}\rightarrow 
p\sp{2}-m\sb{\Delta}\sp{2}+im\sb{\Delta}\Gamma\sb{\Delta}(p\sp{2})$.
The width $\Gamma\sb{\Delta}(s)$ is~\cite{JM97}
\begin{equation}
\Gamma\sb{\Delta}(s)=\Gamma\sb{\Delta}
\frac{m\sb{\Delta}}{\sqrt{s}}
\frac{|{\bf q}(\sqrt{s})|\sp{3}}{|{\bf q}(m\sb{\Delta})|\sp{3}}
\theta(s-(m\sb{N}+m\sb{\pi})\sp{2}),\label{eq:35}
\end{equation}
where ${\bf q}={\bf q}(\sqrt{s})$ is the pion spatial momentum in 
the center of mass $\pi N$ system with the total energy $\sqrt{s}$.

Finally we list the following relations for the isospin matrices 
(see appendix A in Ref. \cite{OV85}),
\begin{eqnarray}
\isomtx{a\dag}{3}{1}\isomtx{b}{3}{1}&=&
\delta\sp{ab}-\frac{1}{3}\tau\sp{a}\tau\sp{b},
\label{eq:36}\\
\isomtx{a\dag}{3}{1}\isomtx{b}{3}{3}\isomtx{c}{3}{1}&=&
\frac{5}{6}i\varepsilon\sp{abc}
-\frac{1}{6}\delta\sp{ab}\tau\sp{c}
+\frac{2}{3}\delta\sp{ac}\tau\sp{b}
-\frac{1}{6}\delta\sp{bc}\tau\sp{a}.\label{eq:37}
\end{eqnarray}
%
%
%
%
%\bibliography{reference}% Produces the bibliography via BibTeX.
%
%

%
%
%
%
\end{document}